\documentclass[11pt,reqno]{amsart}
\usepackage{amssymb}

\usepackage{enumitem}
\usepackage{amsthm,amsfonts,amssymb,euscript,color,bbm}
\usepackage{comment}
\usepackage{mathtools}
\usepackage{stmaryrd}
\usepackage{mathrsfs}
\usepackage{pifont}
\usepackage{amsfonts,amsmath,amssymb,amsthm,amscd,cancel}
\usepackage{graphicx}
\usepackage[]{geometry}
\usepackage{verbatim}
\usepackage{mathrsfs}
\usepackage{fancyhdr}
\usepackage{footnpag,footmisc}
\usepackage[normalem]{ulem}
\usepackage{float}

\usepackage{cite}
\usepackage{hyperref}
\usepackage{engord}
\usepackage[displaymath]{lineno}
\usepackage[utf8]{inputenc}

\theoremstyle{definition}
\newtheorem{theorem}{Theorem}[section]

\newtheorem{proposition}{Proposition}[section]

\newtheorem{remark}{Remark}[section]

\allowdisplaybreaks[4]

\DeclareMathAlphabet{\mathsfsl}{OT1}{cmss}{m}{sl}
\numberwithin{equation}{section}

\newcommand{\D}{\mathrm{d}}

\newcommand{\tr}{\mathrm{tr}}

\def\alphab{\underline{\alpha}}
\def\betab{\underline{\beta}}
\def\chib{\underline{\chi}}
\def\chibh{\widehat{\underline{\chi}}}
\def\chih{\widehat{\chi}}
\def\etab{\underline{\eta}}
\def\Hb{\underline{H}}
\def\Lb{\underline{L}}

\def\tr{\mathrm{tr}}
\def\omegab{\underline{\omega}}

\def\sigmac{\check{\sigma}}
\def\tensor{\widehat{\otimes}}

\def\ub{\underline{u}}

\def\Hb{\underline{H}}

\def\Lh{\widehat{L}}
\def\Lbh{\widehat{\underline{L}}}

\newcommand{\Db}{\underline{D}}
\newcommand{\Dh}{\widehat{D}}
\newcommand{\Dbh}{\widehat{\underline{D}}}

\def\nablas{\mbox{$\nabla \mkern -13mu /$ }}
\def\Deltas{\mbox{$\Delta \mkern -13mu /$ }}

\def\divs{\mbox{$\mathrm{div} \mkern -13mu /$ }}
\def\curls{\mbox{$\mathrm{curl} \mkern -13mu /$ }}

\def\ds{\mbox{$\nabla \mkern -13mu /$ }}
\def\gs{\mbox{$g \mkern -9mu /$}}

\def\epsilons{\mbox{$\epsilon \mkern -9mu /$}}

\def\Ks{\mbox{$K \mkern -13mu / $}}

\begin{document}
	
	\title[Outer trapped surface formation]{Outer trapped surface formation from gravitational waves collision}

	\author[Junbin Li]{Junbin Li}
	\address{Department of Mathematics, Sun Yat-sen University, Guangzhou, China}
	\email{lijunbin@mail.sysu.edu.cn}

	\begin{abstract}
		
	We give a new rigorous construction of outer trapped surface formation in smooth vacuum spacetimes without symmetry assumptions. The outer trapped surface is produced by the collision of two transversely   localized, nearly plane-fronted gravitational waves described by short pulse data.   The outer trapped surface is obtained by joining two smooth pieces along a curve with their future-directed outer null normals matching continuously. The incompleteness of the maximal future development is then established by adapting Penrose's original argument.  This construction is inspired by Penrose's argument in constructing a marginally outer trapped surface in the collision of two ultra-relativistic particles modeled by the Aichelburg--Sexl solution.

	\end{abstract}
	
	\maketitle
	

	\setcounter{tocdepth}{1}


	\section{Introduction}

	In general relativity, trapped surface formation is a central problem in understanding the dynamics of the Einstein equations. With the help of Penrose's incompleteness theorem \cite{Penrose1965}, the problem of trapped surface is a fundamental first step in studying the problem of singularity formation in the Einstein equations, and it is also an essential step in the understanding of the weak cosmic censorship conjecture in view of Christodoulou's trapped surface conjecture \cite{Chr99cqg}.	
	
\begin{figure}
	\includegraphics[width=4in]{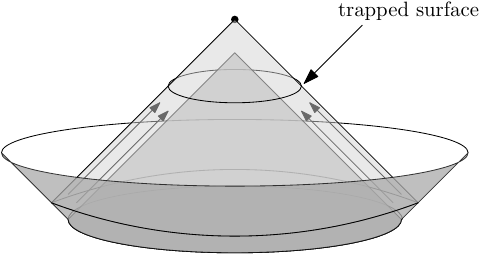}
	\caption{}
	\label{fig:christodoulou}
	\end{figure}

	 In this paper, we study the trapped surface formation for the vacuum Einstein equations
	\begin{equation}\label{EVE}\mathbf{Ric}=0.\end{equation}
The trapped surface (and singularity) formation problem is particularly challenging in vacuum, since Birkhoff theorem rules out nontrivial dynamics in spherical symmetry， and the lack of spherical symmetry introduces substantial additional difficulties. Christodoulou first made a breakthrough \cite{Chr08}, by showing that focusing of gravitational waves can produce closed trapped surfaces (see Figure \ref{fig:christodoulou}). Essentially, his initial data are imposed on two intersecting null cones, with the incoming one being Minkowskian. The data on the outgoing null cone, consisting of the shear tensor $\chih$,  verifies a scaling relation\footnote{In fact, certain estimates compatible with this relation suffice and the exact scaling is not required.}
	$$\chih(\ub, \vartheta)=\delta^{-\frac{1}{2}}\chih_0(\frac{\ub}{\delta},\vartheta),$$
	where $s\in[0,1]$ and $\vartheta\in S^2$. So the genuine data on the initial null cone is defined in a very short interval $\ub\in[0,\delta]$ with a very large amplitude. The gravitational wave described by these data propagates along the incoming null direction and eventually concentrates in a very small region, leading to the formation of a closed trapped surface whose area radius is comparable to the total incoming energy. The main difficulty is to show that the desired a priori estimates remain valid in the region where a closed trapped surface forms. This result has since been extended in various directions \cite{An20, An-Han, An-Luk, K-R12, Luk-Rod17, K-L-R, Li-Yu, Li-Mei, A, Li-Liu1, LiLi, CK}.  However, all these results, including the anisotropic results, rely on the fact that the incoming gravitational waves have contracting spherical wavefronts, with the focusing driven primarily by the contraction of the spherical sections.
	
\begin{figure}
	\includegraphics[width=4in]{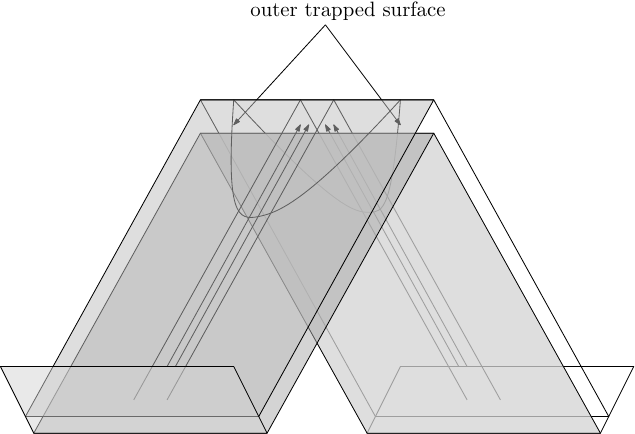}
	\caption{}
	\label{fig:collision}
	\end{figure}

	In this paper we give a new rigorous construction for trapped surface formation in smooth vacuum spacetime, arising from the collision of two nearly plane-fronted gravitational waves, localized in transverse directions, distinct from previous constructions based on the contraction of spherical sections. Nearly planar wavefronts do not undergo significant geometric contraction during propagation. Instead, the collision of two oppositely propagating waves leads to a concentration of gravitational energy sufficient to produce a closed outer trapped surface. The rough version of the theorem is the following:
	\begin{theorem}[Rough version]\label{mainrough}
	There exist smooth, asymptotically flat characteristic initial data for the vacuum Einstein equations, without any symmetries, describing two transversely localized gravitational wave beams propagating toward each other, such that their future development contains a piecewise smooth, closed outer trapped surface with a continuous outer null normal  formed as a result of the collision. Moreover, the maximal future development is future null geodesically incomplete.  The transverse and longitudinal widths of the beams can be made arbitrarily small, while their incoming energies can be prescribed arbitrarily. 
	\end{theorem}
	Here, the transverse directions refer to the two-dimensional planar sections, while the longitudinal directions refer to the null directions. The precise statement is present in Theorem \ref{main}. 
\begin{remark}
In this paper, we only consider gravitational wave beams whose transverse widths are much smaller than the area radius of the trapped surface, with the two beams carrying approximately equal incoming energies. The cases that the transverse widths are large and the beams carry different incoming energies will be treated later. 
\end{remark}
		The construction of such a trapped surface formation spacetime is constructed in the following two steps (see Figure \ref{fig:collision}):
	
{\bf Step 1:} Initial data and solution construction.  Instead of two intersecting null cones with spherical sections, we prescribe characteristic initial data on a W--shaped configuration of null hypersurfaces with planar sections, with Minkowski data on the two inner branches and short-pulse data on the two outer branches. The data is of short pulse type, and localized in transverse directions, representing localized nearly plane-fronted gravitational waves. This problem can also be formulated as a characteristic initial value problem on a V-shaped configuration of null hypersurfaces, with Minkowski data on the inner portions and short-pulse data on the outer portions. Since the data are localized in the transverse directions, they will become asymptotically flat after extending the shear by zero. In future works, we will consider the spacelike Cauchy data formulation, see for example \cite{Li-Yu, Li-Mei, CK}.  For such data,  the solution and the a priori estimates can be obtained by existing results. Note that although we are considering planar sections, but since the data is compactly supported in transverse directions, all connection coefficients and curvature components are compactly supported on each future planer section (note that the background values are all zero in double null-plane foliation). Then previous results can be directly applied, with the Sobolev inequalities and elliptic estimates depending on the size of the support. Before the intersection of two waves, the a priori estimates can be obtained by applying Christodoulou's original argument \cite{Chr08}, or Li-Mei's simplification \cite{Li-Mei} without using elliptic estimates.  For the interaction region, which is of size $\delta\times\delta$, we apply Luk--Rodnianski's result in \cite{Luk-Rod17}, which is the first rigorous mathematical work in the study of the interaction impulsive gravitational waves without symmetry assumptions, and a systematic theory of weak solutions of the vacuum Einstein equations allowing $\delta$-singular Weyl tensor was developed. 

{\bf Step 2:} The outer trapped surface $\mathcal{S}$ is constructed in the way that it is obtained by joining two smooth pieces of outer trapped surface located on the wavefronts the two gravitational waves respectively, along a joining curve $\mathcal{C}$ in the collision section.  These two pieces are viewed as graphs of functions on a ball in the collision section and similar procedure was  done in \cite{K-L-R} in the case of spherical sections.  The surface we construct is clearly not smooth. In order to apply Penrose's incompleteness argument, two outer normals should be joined continuously. Then the outer part of $\partial J^+(\mathcal{S})$ is compact.  The inner null normal is not continuous, and even the inner expansion is not necessarily negative. But we will show that the inner part of $\partial J^+(\mathcal{S})$ is (a closed subset of) the union of the parts of the null hypersurfaces enclosed by the smooth pieces of $\mathcal{S}$ and the joining curve $\mathcal{C}$. Penrose's argument then applies to this case and the resulting maximal future development is future null geodesically incomplete.  Although $\mathcal{S}$ we construct is not a closed trapped surface in a usual sense, but it still predicts the formation of black hole because we still have a Penrose type incompleteness if the weak cosmic censorship conjecture holds true.  There should not be conceptional difficulties in finding a usual smooth closed trapped surface in the spacetime we construct. 

\begin{remark}Note, in particular that a single such localized planar gravitational wave is not expected to  produce a closed trapped surface, for it can be shown that the solution remains regular for any finite time if $\delta$ is chosen small enough by short pulse estimates and the global existence is expected to hold true since the wave can be regarded as an outgoing gravitational wave after completing the planar sections into spheres on the opposite side. This also shows that no closed trapped surface is present initially. 
\end{remark}
	\begin{remark}  This construction can, in principle, be generalized to other types of sections, such as spherical sections, since the background geometry becomes approximately flat upon rescaling to the scale of the localized waves. For example, in Christodoulou's setting, we can prescribe initial data on the incoming null cone to represent an additional outgoing localized gravitational wave. The original wave can also be chosen to be localized in angular directions. A closed trapped surface should then form by a mechanism similar to the one described above. But  the absence of the outgoing wave does not preclude the formation of trapped surface further along the incoming evolution (see also \cite{K-L-R}), owing to the contraction of the spherical sections. In contrast, planar sections allow us to isolate the focusing mechanism arising purely from the collision of the two gravitational waves. But also note that the deformation technique in \cite{K-L-R} is crucial in our construction. 
\end{remark}

	This construction is inspired by Penrose's argument \cite{Penrose1974} in studying collision of two Aichelburg--Sexl solutions \cite{AS}, which can be used to model the spacetime outside a ultra-relativistic particle. Penrose's argument, not published, was presented and generalized by, for example, Eardley--Giddings \cite{EG},  Kohlprath--Veneziano \cite{KV}, and the numerical study of Pretorius--East \cite{PE} which were motivated by producing a black hole by collision of high-energy particles. 
	
 The Aichelburg--Sexl solution takes the form of a pp--wave metric in Brinkmann coordinates
	$$\D s^2=-\D \bar{u}\D \bar{v}+(\D \bar{x}^i)^2+\Phi(\bar{x}^i)\delta(\bar{u})\D\bar{u}^2,$$
	where $\delta(u)$ is the Dirac delta at $u=0$ and
	$$\Phi=-8G\mu\log\bar{\rho},$$
	with $\mu$ represents its center-of-mass energy and $G$ the gravitational constant.  It is the Green function on $\mathbb{R}^2$
	$$\Delta \Phi=-16 \pi G\mu\delta_0(\bar{x}^i).$$
	where $\delta_0$ is the Dirac delta at the origin of $\mathbb R^2$. The only nontrivial components of Ricci tensor and Riemann curvature tensors are 
	$$\mathbf{Ric}_{\bar{u}\bar{u}}=-\frac{1}{2}\Delta\Phi\delta(u)=8\pi G\mu\delta_0\delta(u).$$
	$$\mathbf{R}_{\bar{u}A\bar{u}B}=-\frac{1}{2}\partial_A\partial_B\Phi\delta(u).$$
The Aichelburg--Sexl solution belongs to the class of impulsive pp--waves. A related class of vacuum impulsive plane waves was studied by Penrose \cite{Penrose1972}, whose curvature profiles are homogeneous in the transverse directions and extend over the entire transverse plane. The collision of such impulsive gravitational waves and spacelike singularity formation was studied by, for example  Khan--Penrose \cite{KP},  Szekeres \cite{Szekeres1972} and Yurtsever \cite{Yurtsever1988a} (see \cite{Luk-Rod17} for the references therein).  In contrast, the Aichelburg--Sexl solution describes the gravitational field generated by a single massless particle. Under the coordinate change
$$ \bar{u}=u,$$
$$\bar{v}=v+\Phi\theta(u)+\frac{u\theta(u)|\nabla\Phi|^2}{4},$$
$$\bar{x}^i=x^i+\frac{u}{2}\nabla_i\Phi(x)\theta(u)$$
where $\theta$ is the Heaviside step function, the metric coefficients take a form in a Rosen-type continuous coordinates
$$\D s^2=-\D u\D v+ H_{ik}H_{jk}\D x^i\D x^j, \quad H_{ij}=\delta_{ij}+\frac{u\theta(u)}{2}\partial_i\partial_j\Phi .$$
Away from the particle, the coefficients are locally Lipschitz, with jump discontinuities in their first derivatives, while the curvature contains Dirac delta terms supported on the wavefront. The surface $v=-\Psi(x^i)=-8G\mu\log\frac{4 G\mu}{r}$ is then a marginally outer trapped surface with boundary $\mathcal{C}: \sum(x^i)^2=(4G\mu)^2$ on $v=0$. Two such pieces can be constructed on two head-on colliding Aichelburg--Sexl wavefronts and joined along $\mathcal{C}$ on the collision section. On $\mathcal{C}$, we have $\partial_r\Phi=-2$, which ensures that two such pieces are joined such that the outer null normals match continuously. Note that this marginally outer trapped surface is unbounded in continuous coordinates, but admits a bounded representation in Brinkmann  coordinates (see \cite{EG}, where non-head-on collisions are included). The singular behavior in both the transverse and longitudinal directions poses mathematical difficulties and may raise concerns about the physical applicability of the model. To address these issues, Pretorius--East \cite{PE}  considered collisions of massless Vlasov beams with finite widths in both directions, developing numerical methods to study the resulting black hole formation and gravitational radiation. Earlier, Kohlprath--Veneziano \cite{KV} had extended the construction of Penrose to collisions of null matter beams with finite transverse width, including homogeneous beams and more general axisymmetric profiles, while retaining the impulsive structure in the longitudinal direction. In their setting, $\Phi$ becomes the gravitational potential generated by a continuous matter distribution. This is in turn related to the analysis of finite and large transverse  width beam collision by Yurtsever \cite{Yurtsever1988b}. 

The current work can then be viewed as a realization of Penrose's construction in the setting of smooth solutions of the vacuum Einstein equations without any symmetries.  This construction is also inspired by the mathematical studies of the (reverse) Burnett conjecture \cite{Burnett, HL, LR-HF}, which aim to establish a connection between the high-frequency solutions of the vacuum Einstein equations and the solutions of Einstein--massless Vlasov (including null dust). See for example the survey of Huneau--Luk \cite{HL}. In particular, Luk--Rodnianski \cite{LR-HF} established a correspondence between high-frequency vacuum solutions and Einstein--null dust solutions in double null gauge, thereby providing a new perspective on the relationship between Christodoulou's trapped surface formation \cite{Chr08} and the constructions of Synge--Gibbons--Penrose (see also discussions and references in \cite{LR-HF}). Therefore, Christodoulou's short pulse data provide a natural candidate for realizing Penrose's construction within the framework of smooth vacuum solutions. In view of the arguments of Luk--Rodnianski, the solutions constructed here are expected to converge in a suitable weak sense, to a solution describing outer trapped surface formation from colliding null dust with positive (but small) transverse width. However, whether they converge to Penrose's construction in collision of two Aichelburg--Sexl solutions as the transverse width also tends to zero remains unknown, due to the presence of Dirac delta singularities in the transverse directions. We mention the related work of Li--Li \cite{LiLi}, which discussed trapped surface formation from gravitational waves focusing whose angular support shrinks to a single point.

\subsection*{Acknowledgement}

 The authors are supported by National Key R\&D Program of China (No. 2022YFA1005400) and NSFC (12326602, 12141106).
 
 \subsection*{AI Disclosure}
ChatGPT was used to assist with exploring relevant physics literature, clarifying physical aspects of the problem, checking some calculations, and language editing.

	\section{Double null foliations}
	
	In this section, we introduce the double null foliations. This paper does not contain heavy hyperbolic estimates so only a few equations will be used.  The notations  follow those in Christodoulou's monograph \cite{Chr08}. We assume the solution $(\mathcal{M},g)$ is foliated by two optical functions, $u$ and $\ub$, that is
 $$g(\nabla u,\nabla u)=g(\nabla\ub,\nabla\ub)=0.$$
 In addition, we require that $u$ and $\ub$  increase towards the future. We use $H_u$ to denote the  null hypersurfaces that are the level sets of $u$ and use ${\Hb}_{\ub}$ to denote the incoming null hypersurfaces that are the level sets of $\ub$. We denote the intersection $\Sigma_{\ub,u}=\Hb_{\ub} \cap H_u$, which is a  space-like two-surface.   In this paper, $\Sigma_{\ub,u}$ is topologically $\mathbb{R}^2$. The space-time metric $g$ induces a Riemannian metric $\gs$ on $\Sigma_{\ub,u}$ and $\epsilons$ is the volume form of $\gs$ on $\Sigma_{\ub,u}$. We use   $\nablas$ to denote the  covariant derivative (with respect to $\gs$) on $\Sigma_{\ub,u}$ and $\Ks$ is the Gauss curvature of $\Sigma_{\ub,u}$.

The lapse function $\Omega$ is defined by the formula
$$ \Omega^{-2}=-2g(\nabla\ub,\nabla u).$$
  We then define the normalized null pair $(\Lbh, \Lh)$ by
  $$e_3=\Lbh=-2\Omega\nabla\ub,\ e_4=\Lh=-2\Omega\nabla u.$$
  We have $g(\Lh,\Lbh)=-2$. We also define another null pair
  $$\Lb=\Omega \Lbh,\ L=\Omega \Lh.$$
  We have $g(L,\Lb)=-2\Omega^2$.  The flows generated by $\Lb$ and $L$ preserve the double null foliation. Moreover, we define
 $$\Lb'=-2\nabla\ub=\Omega^{-1}\Lbh,\ \Lb'=-2\nabla u=\Omega^{-1}\Lh$$
 which is geodesic. 

We can introduce the double null coordinate system $(\ub,u,x^A)$ on $\mathcal{M}$, where $A,B,C,\cdots$ are denoted an index from $1$ to $2$. In such a coordinate system, the Lorentzian metric $g$ takes the following form
\begin{align}\label{doublenullmetric}
g=-2\Omega^2(\D\ub\otimes\D u+\D u\otimes\D \ub)+\gs_{AB}(\D x^A-b^A\D u)\otimes(\D x^B-b^B\D u),
\end{align}
where the vectorfield $b=b^A\partial_{x^A}$ is tangent to $\Sigma_{\ub,u}$. The null vectors $\Lb$ and $L$ can be computed as $\Lb=\partial_u+b^A\partial_{x^A}$ and $L=\partial_{\ub}$.


Using the null frame $(e_1,e_2,\Lbh,\Lh)$, the connection coefficients can be decomposed as the following tensorfields on $\Sigma_{\ub,u}$:
\begin{align*}
\chi_{AB}&=g(\nabla_A\Lh,e_B),\quad \eta_A=-\frac{1}{2}g(\nabla_{\Lbh}e_A,\Lh),\quad \omega=\frac{1}{2}\Omega g(\nabla_{\Lh}\Lbh,\Lh),\\
\chib_{AB}&=g(\nabla_A\Lbh,e_B), \quad\etab_A=-\frac{1}{2}g(\nabla_{\Lh}e_A,\Lbh), \quad\omegab=\frac{1}{2}\Omega g(\nabla_{\Lbh}\Lh,\Lbh).
\end{align*}
We also define the following normalized quantities:
$$\chi'=\Omega^{-1}\chi,\ \chib'=\Omega^{-1}\chib,\ \zeta=\frac{1}{2}(\eta-\etab).$$
 The ($2$-dimensional) trace of $\chi$ and $\chib$ are denoted by
 $$\tr\chi = \gs^{AB}\chi_{AB},\ \tr\chib = \gs^{AB}\chib_{AB},$$
 and the trace-free parts of $\chi$ and $\chib$ are denoted by
 $$\chih=\chi-\frac{1}{2}\tr\chi\gs,\ \chibh=\chib-\frac{1}{2}\tr\chib\gs.$$ 
  By definition, we can check directly the following useful identities :
  $$\ds\log\Omega=\frac{1}{2}(\eta+\etab),\ D\log\Omega=\omega,\ \Db\log\Omega=\omegab,$$
and
$$\frac{\partial b^A}{\partial \ub}=4\Omega^2\zeta^\#.$$
So the metric coefficients $\Omega, b^A$ and $\gs_{AB}$ can be recovered by integrating the connection coefficients. 

The curvature tensor $\mathbf{R}$ can be decomposed as
\begin{align*}
\alpha_{AB}&=\mathbf{R}(e_A,\Lh,e_B,\Lh),\quad\beta_A=\frac{1}{2}\mathbf{R}(e_A,\Lh,\Lbh,\Lh),\quad\rho=\frac{1}{4}\mathbf{R}(\Lbh,\Lh,\Lbh,\Lh),\\
\alphab_{AB}&=\mathbf{R}(e_A,\Lbh,e_B,\Lbh),\quad\betab_A=\frac{1}{2}\mathbf{R}(e_A,\Lbh,\Lbh,\Lh),\quad\sigma=\frac{1}{4}\mathbf{R}(\Lbh,\Lh,e_A,e_B)\epsilons^{AB}.
\end{align*}

    Under the assumption that the spacetime verifies the vacuum Einstein equations \eqref{EVE}, we have the followings are the \textit{null structure equations}, where $D$ and $\Db$ is the projection to $\Sigma_{\ub,u}$ of the usual Lie derivatives $\mathcal{L}_L$ and $\mathcal{L}_{\Lb}$ respectively:
        \begin{align*}
    \Db(\Omega\tr\chib)&=-\frac{1}{2}(\Omega\tr\chib)^2+2\omegab\Omega\tr\chib-|\Omega\chibh|^2,\\
    D\tr\chi'&=-\frac{1}{2}(\Omega\tr\chi')^2-|\chih|^2,\\
    \Dbh(\Omega\chih)&=\Omega^2(\nablas \tensor \eta + \eta \tensor \eta +\frac{1}{2}\tr\chib\chih-\frac{1}{2}\tr\chi \chibh),\\
    \Dh(\Omega\chibh)&=\Omega^2(\nablas \tensor \etab + \etab \tensor \etab +\frac{1}{2}\tr\chi\chibh-\frac{1}{2}\tr\chib \chih),\\
    \Db(\Omega\tr\chi)&=\Omega^2(2\divs\eta+2|\eta|^2-\tr\chi\tr\chib-2\Ks),\\
    D(\Omega\tr\chib)&=\Omega^2(2\divs\etab+2|\etab|^2-\tr\chi\tr\chib-2\Ks),\\
    D\eta &= (\Omega\chi)\cdot\etab-\Omega\beta,\\
    \Db\etab &= (\Omega\chib) \cdot\eta+\Omega\betab,\\
    D  \omegab &=\Omega^2(2(\eta,\etab)-|\eta|^2-\rho),\\
    \Db  \omega &=\Omega^2(2(\eta,\etab)-|\etab|^2-\rho),\\
    -\rho&=\Ks+\frac{1}{4}\tr \chi\tr\chib-\frac{1}{2}(\chih,\chibh),\\
    \divs(\Omega\chih)&=\frac{1}{2}\Omega^2\ds\tr \chi'+\Omega\chih\cdot\etab+\frac{1}{2}\Omega\tr \chi\eta-\Omega\beta,\\
    \divs(\Omega\chibh)&=\frac{1}{2}\Omega^2\ds\tr \chib'+\Omega\chibh\cdot\eta+\frac{1}{2}\Omega\tr \chib\etab+\Omega\betab.\\
    \end{align*}

	The contracted second Bianchi identity will accordingly be the following:
	\begin{equation*}
		\nabla^{\alpha}\mathbf{R}_{\alpha\beta\gamma\delta}=0.
	\end{equation*}
	This equation can be decomposed using the null frame $(e_1,e_2,e_3,e_4)$ into components, which we call \textit{null Bianchi equations}:

    \begin{align*}
    &\Dbh\alpha-\frac{1}{2}\Omega\tr\chib \alpha+2\omegab\alpha+\Omega\{-\nablas\tensor\beta -(4\eta+\zeta)\tensor \beta+3\chih \rho+3{}^*\chih \sigma\}=0,\\
    &D\beta+\frac{3}{2}\Omega\tr\chi\beta-\Omega\chih\cdot\beta-\omega\beta-\Omega\{\divs\alpha+(\etab+2\zeta)\cdot\alpha\}=0,\\
    &\Db\beta+\frac{1}{2}\Omega\tr\chib\beta-\Omega\chibh \cdot \beta+\omegab \beta-\Omega\{\ds \rho+{}^*\ds \sigma+3\eta\rho+3{}^*\eta\sigma+2\chih\cdot\betab\}=0,\\
    &D\rho+\frac{3}{2}\Omega\tr\chi \rho-\Omega\{\divs \beta+(2\etab+\zeta,\beta)-\frac{1}{2}(\chibh,\alpha)\}=0,\\
    &D\sigma+\frac{3}{2}\Omega\tr\chi\sigma+\Omega\{\curls\beta+(2\etab+\zeta,{}^*\beta)-\frac{1}{2}\chibh\wedge\alpha\}=0,\\
    &D\betab+\frac{1}{2}\Omega\tr\chi\betab-\Omega\chih \cdot \betab+\omega \betab+\Omega\{\ds \rho-{}^*\ds \sigma+3\etab\rho-3{}^*\etab\sigma-2\chibh\cdot\beta\}=0,\\
    &\Db\rho+\frac{3}{2}\Omega\tr\chib \rho+\Omega\{\divs \betab+(2\eta-\zeta,\betab)+\frac{1}{2}(\chih,\alphab)\}=0,\\
    &\Db\sigma+\frac{3}{2}\Omega\tr\chib\sigma+\Omega\{\curls\betab+(2\eta-\zeta,{}^*\betab)+\frac{1}{2}\chih\wedge\alphab\}=0, \\
    &\Db\betab+\frac{3}{2}\Omega\tr\chib\betab-\Omega\chibh\cdot\betab-\omegab\betab+\Omega\{\divs\alphab+(\eta-2\zeta)\cdot\alphab\}=0, \\
    &\Dh\alphab-\frac{1}{2}\Omega\tr\chi \alphab+2\omega\alphab+\Omega\{\nablas\tensor\betab +(4\etab-\zeta)\tensor \betab+3\chibh \rho-3{}^*\chibh \sigma\}=0. \\
     \end{align*}
In these equations, the hat sign denotes the trace-free parts of corresponding tensors and star sign denotes the Hodge dual on $\Sigma_{\ub,u}$.

\section{The initial value problem on two intersecting null planes}

\begin{figure}
	\includegraphics[width=3.6in]{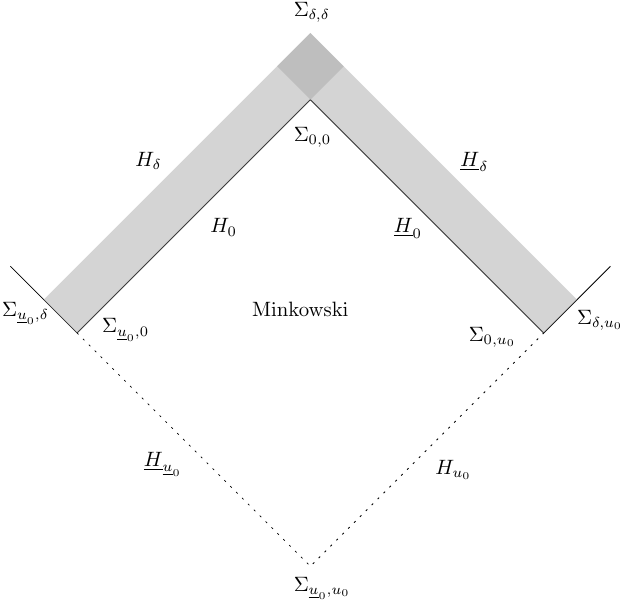}
	\caption{}
	\label{fig:solution}
	\end{figure}

The initial data considered in this paper is given in the following way (see Figure \ref{fig:solution}): The data are given on two intersecting null hypersurfaces, say $H_{\ub_0}$ and $\Hb_{u_0}$, where $\ub_0=u_0$ are given negative constants. There intersection is $\Sigma_{\ub_0,u_0}$, which is a flat $\mathbb{R}^2$. The data for $\ub\in [\ub_0, 0]$ and $u\in [u_0,0]$ is identical to the Minkowski intersecting null planes. Hence the solution of the vacuum Einstein equations is simply Minkowski space for $(\ub,u)\in [\ub_0, 0]\times [u_0,0]$.  Suppose that the remaining two coordinates $(x,y)$ is determined by the condition that they are the standard coordinates restricted on $\Sigma_{-\ub_0,u_0}$. The standard Cartesian coordinates $(t,x,y,z)$ on Minkowski space are related to these coordinates by
$$2\ub=t+z, \quad 2u=t-z,$$
in which the Minkowski metric reads
$$-4\D \ub\D u+ \D x^2+\D y^2.$$
We denote 
$$T=-2\ub_0=-2u_0$$
the approximate time to collision in Minkowski coordinates.

The nontrivial data is the conformal metrics on $H_{-\ub_0}$ for $\ub\in[0,\delta]$ and on $\Hb_{u_0}$ for $u\in [0,\delta]$.  We take a matrix value function that is symmetric and trace-free on $[0,1]\times\mathbb{R}^2$, called seed data:
$$\psi(s, X,Y)$$
whose support in $(X, Y)$ is in $\{X^2+Y^2\le 1\}$ and it admits a smooth zero extension to $s<0$ and $s>1$. 
We then define the conformal metric in this coordinate system $(\ub,u,x,y)$ to be
$$\widehat{\gs}(\ub, u_0, x,y)=\exp(\delta^{\frac{1}{2}}\epsilon^{-1}\psi (\frac{\ub}{\delta}, \frac{x}{\epsilon},\frac{y}{\epsilon})),$$
$$\widehat{\gs}(\ub_0, u, x,y)=\exp(\delta^{\frac{1}{2}}\epsilon^{-1}\psi (\frac{u}{\delta},\frac{x}{\epsilon},\frac{y}{\epsilon})).$$
Here $\epsilon>0$ is a positive number, denoting the transverse beam width.  The above matrices are then of determinant $1$ and define the initial  conformal metrics (see more discussions in \cite{Chr08}). By zero extension, we have data on two $H_{\ub_0}\cup \Hb_{u_0}$ which is asymptotically flat as in \cite{Chr08}. In applying Penrose type incompleteness argument, the non-compactness of the initial data suffices.   Of course one can use two different $\psi_1,\psi_2$ to define two conformal metrics.  

We assume that the seed data is chosen such that
$$\int_0^1|\partial_s\psi(s,X,Y)|^2\D s>0,\quad X^2+Y^2\le 0.75,$$
then for some $M_0>0$
\begin{equation}\label{incomingenergy}
\begin{split}
\int_0^\delta|\chih(\ub, u_0, x,y)|^2\D\ub \ge M_0\epsilon^{-2}, \quad x^2+y^2\le 0.75\epsilon^2,\\
\int_0^\delta|\chibh(\ub_0, u, x,y)|^2\D u\ge M_0\epsilon^{-2}, \quad x^2+y^2\le 0.75\epsilon^2,
\end{split}
\end{equation}
Therefore the totally incoming energy is at least $M_0$ for each beam.

We are going to present the precise statement of Theorem \ref{mainrough}.
\begin{theorem}[Precise version of the main result]\label{main}
Consider the characteristic initial value problem as above. $M_0,\epsilon$ are arbitrarily chosen such that $M_0\gg\epsilon>0$ and $T>0$ is chosen such that $T\gg\log\frac{1}{\epsilon}$.  If $\delta$ is sufficiently small, the the solution of the vacuum Einstein equations remain smooth in double null coordinates $(\ub,u)\in [\ub_0,\delta]\times[u_0,\delta]$. A closed, piecewise smooth outer trapped surface forms in this region, with its outer null normal being continuous. The maximal future development of such initial data is future null geodesically incomplete. The area radius of the outer trapped surface is comparable to $M_0$. 
\end{theorem}

Now we consider the full initial data set on $H_{u_0}$ and the data on $\Hb_{\ub_0}$ is similar.  By similar argument as in \cite{Chr08} we have 
$$|\chih(\ub,u_0, x,y)|\lesssim \delta^{-\frac{1}{2}}$$
on $H_{u_0}$ for $\ub\in[0,\delta]$.  The implicit constant depends on $\psi$ (and hence on $M_0$ and $\epsilon$) and $T$. This estimate is also true for taking transverse derivatives and the constant depends on the order of derivatives. Moreover, the full metric verifies
$$|\gs(\ub,u_0,x,y)-\gs_{\mathrm{Euc}}|\lesssim \delta^{\frac{1}{2}}, \ub\in[0,\delta]$$
where $\gs_{\mathrm{Euc}}$ is the $2$-dimensional Euclidean metric. As a consequence, we have
$$|\Ks(\ub,u_0,x,y)|\lesssim\delta^{\frac{1}{2}},  \ub\in[0,\delta]$$
By integrating the Raychauduri equation along $\ub$ direction, we have
$$|\tr\chi(\ub, u_0,x,y)|\lesssim 1, \ub\in[0,\delta]$$
Note here that the initial value of $\tr\chi$ is zero, as we are working on null planes. The other connection coefficients verify 
$$|\eta|, |\etab|, |\omegab|, |\chibh|, |\tr\chib|\lesssim\delta^{\frac{1}{2}}$$
and the curvature components verify
$$|\alpha|\lesssim\delta^{-\frac{3}{2}}, |\beta|\lesssim\delta^{-\frac{1}{2}}, |\rho|,|\sigma|\lesssim1, |\betab|\lesssim\delta ,|\alphab|\lesssim\delta^{\frac{3}{2}}$$
on $H_{u_0}$ for $\ub\in[0,\delta]$.  We may choose $\Omega\equiv 1$ on $H_{u_0}$ and hence $\omega=0$.  The above all bounds are understood in $C^m$ in transverse directions and the implicit constants depend in addition on $m$. Note that these norms are the same to Christodoulou's work \cite{Chr08}, with the only difference being that we are currently working on nearly planar sections. We will have the following estimates in the regions $(\ub,u)\in[0,\delta]\times [u_0,0]$ and $(\ub,u)\in [\ub_0,0]\times [0,\delta]$. 

\begin{proposition}\label{thm:beforeintersection}
If $\delta$ is sufficiently small depending on $\psi$ and $T$, the solution of the vacuum Einstein equations remains smooth in double null coordinates 
$$(\ub,u)\in[0,\delta]\times [u_0,0]$$
with the following estimates hold
$$\|\Omega-1\|_{H^5(\Sigma_{\ub,u})}, \|b\|_{H^5(\Sigma_{\ub,u})}, \|\gs-\gs_{\mathrm{Euc}}\|_{H^5(\Sigma_{\ub,u})}\lesssim\delta^{\frac{1}{2}},$$
$$\|\chih\|_{H^5(\Sigma_{\ub,u})}\lesssim\delta^{-\frac{1}{2}}, \|\tr\chi, \omega\|_{H^5(\Sigma_{\ub,u})}\lesssim 1, \|\chibh,\tr\chib,\eta,\etab,\omegab\|_{H^5(\Sigma_{\ub,u})}\lesssim\delta^{\frac{1}{2}},$$
$$\|\alpha\|_{L^2_{\ub}H^5_\Sigma(H_{u})}+\|\beta\|_{L^2_uH^5_\Sigma(\Hb_{\ub})}\lesssim \delta^{-1},$$
$$\|\beta\|_{L^2_{\ub}H^5_\Sigma(H_{u})}+\|\rho,\sigma\|_{L^2_uH^5_\Sigma(\Hb_{\ub})}\lesssim 1,$$
$$\|\rho,\sigma\|_{L^2_{\ub}H^5_\Sigma(H_{u})}+\|\betab\|_{L^2_uH^5_\Sigma(\Hb_{\ub})}\lesssim \delta^{\frac{1}{2}},$$
$$\|\betab\|_{L^2_{\ub}H^5_\Sigma(H_{u})}+\|\alphab\|_{L^2_uH^5_\Sigma(\Hb_{\ub})}\lesssim \delta^{\frac{1}{2}},$$
 and in the region
 $$(\ub,u)\in [\ub_0,0]\times [0,\delta]$$
 with the following conjugated estimates hold
$$\|\Omega-1\|_{H^5(\Sigma_{\ub,u})}, \|b\|_{H^5(\Sigma_{\ub,u})}, \|\gs-\gs_{\mathrm{Euc}}\|_{H^5(\Sigma_{\ub,u})}\lesssim\delta^{\frac{1}{2}},$$
$$\|\chibh\|_{H^5(\Sigma_{\ub,u})}\lesssim\delta^{-\frac{1}{2}}, \|\tr\chib, \omegab\|_{H^5(\Sigma_{\ub,u})}\lesssim 1, \|\chih,\tr\chi,\eta,\etab,\omega\|_{H^5(\Sigma_{\ub,u})}\lesssim\delta^{\frac{1}{2}},$$
$$\|\alphab\|_{L^2_uH^5_\Sigma(\Hb_{\ub})}+\|\betab\|_{L^2_{\ub}H^5_\Sigma(H_{u})}\lesssim \delta^{-1},$$
$$\|\betab\|_{L^2_uH^5_\Sigma(\Hb_{\ub})}+\|\rho,\sigma\|_{L^2_{\ub}H^5_\Sigma(H_{u})}\lesssim 1,$$
$$\|\rho,\sigma\|_{L^2_uH^5_\Sigma(\Hb_{\ub})}+\|\beta\|_{L^2_{\ub}H^5_\Sigma(H_{u})}\lesssim \delta^{\frac{1}{2}},$$
$$\|\beta\|_{L^2_uH^5_\Sigma(\Hb_{\ub})}+\|\alpha\|_{L^2_{\ub}H^5_\Sigma(H_{u})}\lesssim \delta^{\frac{1}{2}}.$$
All estimates hold with $H^5_\Sigma$ replaced by $C^3_\Sigma$.
\end{proposition}
The estimates are slightly different from \cite{Chr08}. For example, in the region $(\ub,u)\in[0,\delta]\times [u_0,0]$, $\alphab$ is estimated by $\delta^{\frac{1}{2}}$ here, instead of $\delta^{\frac{3}{2}}$ as in \cite{Chr08}. This is exactly the estimates established in Li--Mei \cite{Li-Mei} in the case for spherical  sections.  So this theorem follows immediately as argued in \cite{Li-Mei}, without using elliptic estimates for the Hodge system, which can be viewed as a simplification of Christodoulou's original argument. 

The only difference is that in the current paper we work on non-compact planar sections. Note that the initial data $\psi$ is compactly supported in an $\epsilon$-ball in $(x,y)$-plane, so for each $\ub\in[0,\delta], u\in [u_0,0]$, all connection coefficients and curvature components has compact supports on $\Sigma_{\ub,u}$. Suppose that the size (in coordinate $(x,y)$) of the support is $R=R(\ub,u)$. The constants of the Sobolev inequalities needed in the a priori estimates now depend on $R$.  In fact, each reduction in derivative order comes at the cost of a factor of $R$, so the implicit constants in the above estimates depend polynomially on $R$.  In this paper we only need to know that $\sup_{\ub,u}R$ is finite and increases as $T$ increases.  The small parameter $\delta$ and the implicit constants are allowed to depend on $\sup_{\ub,u}R$.   By considering the intersection of an expanding null cone and the null plane, which is a parabola in Minkowski,  it is not difficult to see $R\lesssim \sqrt{T}$ so the dependence of the constant on $R$ can be absorbed into its dependence on $T$.

  The next step is to solve the solution in the region $(\ub,u)\in [0,\delta]\times[0,\delta]$, the intersecting region of the gravitational waves.  The data on $H_0\cup \Hb_0$ is induced from the previous solutions and obey a very bad estimate. For example, both of $\chih$ and $\chibh$ can only be bounded in $\delta^{-\frac{1}{2}}$.  But the estimates can still be obtained in view of the work \cite{Luk-Rod17} of Luk--Rodnianski, with both $\chih$ and $\chibh$ being only in $L^2$ initially. We apply Theorem 4 in \cite{Luk-Rod17} to obtain the following.
\begin{proposition}\label{thm:intersectingregion}
If $\delta$ is sufficiently small depending on $\psi$ and $T$, the solution of the vacuum Einstein equations remains smooth in the region 
$$(\ub,u)\in[0,\delta]\times [0,\delta]$$
with the following estimates hold
$$\|\chih, \omega\|_{L^2_{\ub}H^5_\Sigma(H_{u})}\lesssim 1, \|\chibh,\omegab\|_{L^2_uH^5_\Sigma(\Hb_{\ub})}\lesssim 1, $$
$$\| \tr\chi,  \tr\chib,  \eta,\etab\|_{H^5(\Sigma_{\ub,u})}\lesssim 1,$$
$$\|\chih, \omega\|_{L^2_{\ub}C^1_\Sigma(H_{u})}\lesssim 1, \|\chibh,\omegab\|_{L^2_u C^3_\Sigma(\Hb_{\ub})}\lesssim 1, $$
$$\| \tr\chi,   \tr\chib,   \eta,\etab\|_{C^3(\Sigma_{\ub,u})}\lesssim 1,$$
$$\|\nablas^6\chih, \nablas^6\omega, \nablas^6\eta,\nablas^6\etab\|_{L^2_{\ub}L^2_\Sigma(H_{u})}, \|\nablas^6\tr\chi\|_{L^2(\Sigma_{\ub,u})}\lesssim 1$$
$$\|\nablas^6\chibh, \nablas^6\omegab, \nablas^6\eta,\nablas^6\etab\|_{L^2_uL^2_\Sigma(\Hb_{\ub})}\lesssim 1, \|\nablas^6\tr\chib\|_{L^2(\Sigma_{\ub,u})}\lesssim 1$$
$$\|\beta\|_{L^2_{\ub}H^5_\Sigma(H_{u})}+\|\Ks,\sigmac\|_{L^2_uH^5_\Sigma(\Hb_{\ub})}\lesssim 1,$$
$$\|\Ks,\sigmac\|_{L^2_{\ub}H^5_\Sigma(H_{u})}+\|\betab\|_{L^2_uH^5_\Sigma(\Hb_{\ub})}\lesssim 1.$$
\end{proposition}
Only two derivatives of the curvature components are derived in  \cite{Luk-Rod17}, but there are no essential difficulties in obtaining more derivatives. Also note that the estimates  in  \cite{Luk-Rod17} are obtained in the case of spherical sections, which is also applicable for planar sections with compactly supported data as discussed above.
\begin{remark}\label{rmk:metricclosetoMinkowski}
We have additional smallness as compared to the estimates in \cite{Luk-Rod17} because the data on $H_0\cup\Hb_0$ behave better. For example, by directly integrating the metric coefficients from $\Hb_0$, we have
$$\|\Omega-1\|_{C^3(\Sigma_{\ub,u})}, \|b\|_{C^3(\Sigma_{\ub,u})}, \|\gs-\gs_{\mathrm{Euc}}\|_{C^3(\Sigma_{\ub,u})}\lesssim\delta^{\frac{1}{2}}.$$
This means that the metric and its transverse derivatives remain close to their Minkowski counterparts, while its derivatives in the null directions may be large.
\end{remark}
	
	\section{Improved estimates outside the waves}
	
	In this section we established improved estimates for the connection coefficients in the intersecting region $(\ub,u)\times[0,\delta]\times[0,\delta]$ with $(x,y)$ being located outside the $\sqrt{2}\epsilon$-ball transported by null generators. 
	\begin{proposition}\label{improvedoutsidewave}
	For $(\ub,u)\times[0,\delta]\times[0,\delta]$ and $x^2+y^2\ge2\epsilon^2$, if $\delta$ is sufficiently small, we have the following $L^\infty$ estimates for all connection coefficients
	$$|\chih|, |\chibh|, |\tr\chi|, |\tr\chib|, |\eta|, |\etab|, |\omega|, |\omegab|\lesssim\delta^{\frac{1}{2}}.$$
	\end{proposition}
	\begin{proof}
	The basic strategy is integrating the null structure equations of these quantities along null generators initiated outside the ball $x^2+y^2\ge\epsilon^2$ on $H_{u_0}$ and $\Hb_{\ub_0}$. The data is trivial there by construction. This idea was first employed in \cite{Li-Yu} in showing that after the short pulse region of Christodoulou \cite{Chr08} the spacetime becomes close to Schwarzschild with an additional isotropic condition on the initial data. We also need to improve some of the estimates in Proposition \ref{thm:intersectingregion} due to the smallness of the data on $H_0\cup\Hb_0$. 
	
	We first derive improved estimates before intersection. This is exactly the same as in \cite{Li-Yu}. In the region $(\ub,u)\in[0,\delta]\times[u_0,0]$, we only need to improve $\chih$, $\tr\chi$ and $\omega$. For $\chi$ and $\tr\chi$, we rely on the following equations
	  \begin{equation}\label{Dbchih}
	   \Dbh(\Omega\chih)=\Omega^2(\nablas \tensor \eta + \eta \tensor \eta +\frac{1}{2}\tr\chib\chih-\frac{1}{2}\tr\chi \chibh),
	   \end{equation}
	   \begin{equation}\label{Dbtrchi}
    \Db(\Omega\tr\chi)=\Omega^2(2\divs\eta+2|\eta|^2-(\chih,\chibh)-\frac{1}{2}\tr\chi\tr\chib+2\rho).
	\end{equation}
	By the estimates in Proposition \ref{thm:beforeintersection}, the right hand side is already bounded by $\delta^{\frac{1}{2}}$ even inside the waves, except the term $\tr\chib\chih$, which can however be absorbed by the left hand side.  So we know that in the region  $(\ub,u)\in[0,\delta]\times[u_0,0]$, 
	$$|\nablas^{\le2}\chih(\ub,u, x,y)|, |\nablas^{\le2}\tr\chi(\ub,u,x,y)|\lesssim\delta^{\frac{1}{2}},$$
	whenever the null generator of $\Lb$ terminating  at $(\ub,u, x,y)$ starts from a point $(\ub, u_0, \bar{x},\bar{y})$ with $\bar{x}^2+\bar{y}^2\ge\epsilon^2$. The variation of $x$ and $y$ is obtained by the following equation
	$$\Lb x=b^x, \Lb y=b^y.$$
	So the variations of $x$ and $y$ do not exceed $\delta^{\frac{1}{2}}$ and hence the improved estimates hold at least for $x^2+y^2\ge 1.5\epsilon^2$.  From the equation
	$$\divs(\Omega\chih)=\frac{1}{2}\Omega^2\ds\tr \chi'+\Omega\chih\cdot\etab+\frac{1}{2}\Omega\tr \chi\eta-\Omega\beta,$$
	The improved estimates for $\chih$ and $\tr\chi$ give
	$$|\nablas^{\le 1}\beta|\lesssim\delta^{\frac{1}{2}}$$
	for $x^2+y^2\ge 1.5\epsilon^2$. Then integrating the following equation
	$$\Dbh\alpha-\frac{1}{2}\Omega\tr\chib \alpha+2\omegab\alpha+\Omega\{-\nablas\tensor\beta -(4\eta+\zeta)\tensor \beta+3\chih \rho+3{}^*\chih \sigma\}=0$$
	gives
	$$|\alpha(\ub,u, x,y)|\lesssim\delta^{\frac{1}{2}}$$
	 for $x^2+y^2\ge 1.5\epsilon^2$.	With these improved estimates,  we may improve $\rho$ by the equation
	$$D\rho+\frac{3}{2}\Omega\tr\chi \rho-\Omega\{\divs \beta+(2\etab+\zeta,\beta)-\frac{1}{2}(\chibh,\alpha)\}=0,$$
	and finally we obtain improved bound for $\omega$ for $x^2+y^2\ge 1.5\epsilon^2$
	$$|\omega(\ub,u, x,y)|\lesssim\delta^{\frac{1}{2}}. $$
	via the equation
	    \begin{equation}\label{Dbomega}  \Db  \omega =\Omega^2(2(\eta,\etab)-|\etab|^2-\rho).\end{equation}

	The same argument works for the other wave and hence we also have 
	$$|\chibh(\ub,u, x,y)|, |\tr\chib(\ub,u,x,y)|, |\omegab(\ub,u,x,y)|\lesssim\delta^{\frac{1}{2}},$$
	in the region  $(\ub,u)\in[\ub_0,0]\times[0,\delta]$ for $x^2+y^2\ge \epsilon^2$. Note that the coefficient of $\epsilon^2$ here is $1$ because $x,y$ is transported along $L$ so the $\epsilon$-ball in $(x,y)$ coordinates is also transported.

	Now we derive improved estimates in the intersecting region.  We first show that $\eta$ and $\etab$ obey better estimates even inside the waves. This can be seen from the following two equations
\begin{align*}	
 D\eta &= (\Omega\chi)\cdot\etab-\Omega\beta,\\
    \Db\etab &= (\Omega\chib) \cdot\eta+\Omega\betab.\\
\end{align*}
	Note that $\|\eta\|_{H^5(\Sigma_{0,u})}\lesssim\delta^{\frac{1}{2}}$ for $u\in [0,\delta]$. Integrating the first equation, employing the estimates in Proposition \ref{thm:intersectingregion} for $\chih,\tr\chi$ and $\beta$ gives the improved bound
	$$\|\eta\|_{H^5(\Sigma_{\ub,u})}\lesssim\delta^{\frac{1}{2}}.$$
	for $(\ub,u)\in[0,\delta]\times[0,\delta]$. Similarly in the same region we have
	$$\|\etab\|_{H^5(\Sigma_{\ub,u})}\lesssim\delta^{\frac{1}{2}}.$$
	By Sobolev inequality we have $C^3$ improved bounds. Again by \eqref{Dbchih}, together with the improved estimate of $\chih$ at on $H_0$ for $x^2+y^2\ge1.5\epsilon^2$ derived above, we have 
	$$|\nablas^{\le2}\chih(\ub,u,x,y)|\lesssim\delta^{\frac{1}{2}}$$
	for $(\ub,u)\in[0,\delta]\times[0,\delta]$ and $x^2+y^2\ge 2\epsilon^2$. Note that we have an additional $\delta$ factor when integrating along $u\in[0,\delta]$. By \eqref{Dbtrchi}, with this improved bound for $\chih$, we have
	$$|\nablas^{\le2}\tr\chi(\ub,u,x,y)|\lesssim\delta^{\frac{1}{2}}$$
	for $(\ub,u)\in[0,\delta]\times[0,\delta]$ and $x^2+y^2\ge 2\epsilon^2$. By integrating \eqref{Dbomega}, we immediately have
		$$|\nablas^{\le3}\omega(\ub,u,x,y)|\lesssim\delta^{\frac{1}{2}}$$
	for $(\ub,u)\in[0,\delta]\times[0,\delta]$ and $x^2+y^2\ge 2\epsilon^2$.
	 The improved estimates for $\chibh$, $\tr\chib$ and $\omegab$ can be derived similarly. The proof is then completed.
	\end{proof}

	\section{Construction of outer trapped surface and incompleteness of the maximal future development}
	
	In this section complete the proof of Theorem \ref{main} by constructing the outer trapped surface and show that the maximal future development is null geodesically incomplete.  The outer trapped surface consists of two pieces constructed on   $\Hb_\delta$ and $H_\delta$, which is joined along a curve $\mathcal{C}$ on $\Sigma_{\delta,\delta}$ with continuous outer null normal. 
	
	We take 
	$$\mathcal{C}=\{(\delta,\delta, x,y)| x^2+y^2=r_c^2\}\subset\Sigma_{\delta,\delta}$$
	where $r_c$ is to be determined. 
	 The surface on $\Hb_\delta$ is described as a graph of a nonnegative function $\Psi$ defined on the coordinate ball with radius $r_c$
	 $$\underline{S}=\{(\delta, -\Psi(x,y)+\delta, \bar{x}_\Psi(x,y) ,\bar{y}_\Psi(x,y))| x^2+y^2\le r_c^2\},$$
	 with the property
	 \begin{equation}\label{boundarycondition} \Psi(x,y)=0,  \quad x^2+y^2=r_c^2,\end{equation}
	 where $(\delta, -\Psi(x,y)+\delta, \bar{x}_\Psi(x,y) ,\bar{y}_\Psi(x,y))$ is the starting point of the null generator of $\Lb$  terminating at $(\delta,\delta,x,y)$\footnote{This is equivalent to introducing a new coordinate system $(\ub,u,\bar{x},\bar{y})$ so that $\Lb(\bar{x})=\Lb(\bar{y})=0$ on $\Hb_\delta$ and then write $\underline{S}$ as $\{(\delta,-\Psi(\bar{x},\bar{y})+\delta,\bar{x},\bar{y})|\bar{x}^2+\bar{y}^2\le r_c^2\}$.}. 
	 The surface on $H_\delta$ is described by the graph of the same function
	 $$S=\{(-\Psi(x,y)+\delta,\delta, x,y)| x^2+y^2\le r_c^2\},$$
	 with the same boundary condition \eqref{boundarycondition}. Note here that $x,y$ does not change along the generator $L$. Using the same function is merely for convenience and we may also choose different functions on $H_\delta$ and $\Hb_\delta$ especially when one may wish to choose different seed data.

	 Consider the graph $\underline{S}$ on $\Hb_\delta$. We have the following formula for the outer expansion (see for example \cite{K-L-R}, note that the definition of $\omega$ and $\omegab$ differs a little bit)
	$$\overline{\tr\chi}=\tr\chi-2\Omega\Deltas\Psi-4\Omega\eta\cdot\nablas \Psi-4\Omega^2\chibh_{bc}\nablas^b\Psi\nablas^c\Psi-\Omega^2\tr\chib|\nablas \Psi|^2+4\Omega\omegab|\nablas\Psi|^2.$$
which is computed along $\underline{S}$. The function $\Psi$ should be constructed such that 
$$\label{overlinetrchi<0}\overline{\tr\chi}<0,$$
that is, 
\begin{equation}\label{overlinetrchi<0}2\Deltas\Psi-\Omega^{-1}\tr\chi+4\eta\cdot\nablas\Psi+4\Omega\chibh_{bc}\nablas^b\Psi\nablas^c\Psi+\Omega\tr\chib|\nablas \Psi|^2-4\omegab|\nablas \Psi|^2>0\end{equation}
 The graph $S$ of the same function on $H_\delta$ will then be such that $\overline{\tr\chib}<0$ by conjugation.  Now $S$ and $\underline{S}$ are joined continuously along $\mathcal{C}$ but not in $C^1$.  Apparently, the incoming null normals are simply the generators of $\Hb_\delta$ and $H_\delta$, which cannot be joined continuous along at $\Sigma_{\delta,\delta}$.  But it is possible to choose $\Psi$ such that $S$ and $\underline{S}$ are joined such that the outer normals are continuous.  The outer normal on $\underline{S}$ is given by
$$\widehat{\ell}=\widehat{L}-2\Omega\nablas\Psi+\Omega^2|\nablas\Psi| ^2\widehat{\Lb}$$
and the outer normal on $S$ is given by
	$$\widehat{\underline{\ell}}=\widehat{\Lb}-2\Omega\nablas\Psi+\Omega^2|\nablas\Psi|^2\widehat{L}.$$
These two normals must be parallel along $\mathcal{C}$.  The joining condition is then simply
\begin{equation}\label{joiningcondition}\Omega |\nablas\Psi|_{\mathcal{C}}=1.\end{equation}

Now we consider the model solution for $0<r^2=x^2+y^2\le r_c^2$,
$$\Psi_{A,\varepsilon}=\Psi_{A,\varepsilon}(r)=A\log\frac{r_c}{r}+\frac{\varepsilon}{4}(r^2-r_c^2)$$
where $r^2=x^2+y^2$. It satisfies
$$\Delta\Psi_{A,\varepsilon}=\varepsilon-2\pi A\delta_0$$
where $\Delta$ is the Laplacian in flat $\mathbb{R}^2$ and $\delta_0$ is the Dirac delta at $(x,y)=(0,0)$.  $A$ and $\varepsilon$ (which should not be confused with $\epsilon$) are to be determined. We choose $\Psi$ to be
$$\Psi(x,y)=\begin{cases}\Psi_{A,\varepsilon}(r), & x^2+y^2\ge 0.5\epsilon^2\\ \text{smooth cutoff}, & 0.25\epsilon^2\le x^2+y^2\le 0.5\epsilon^2 \\ \Psi_{A,\varepsilon}(0.5\epsilon), &x^2+y^2\le0.25\epsilon^2.\end{cases}$$
We will have 
$$\begin{cases}|\Psi|=O(1+\log \frac{1}{r}), &x^2+y^2\ge 0.25\epsilon^2\\
|\Psi|=O(\log\frac{1}{\epsilon}), \\\nabla \Psi=O(r^{-1}), &x^2+y^2\ge 0.5\epsilon^2\\
\nabla^2\Psi=O(r^{-2}),  &x^2+y^2\ge 0.5\epsilon^2\\
\nabla\Psi=O(\epsilon^{-1}) & 0.25\epsilon^2\le x^2+y^2\le 0.5\epsilon^2 \\
 \nabla^2\Psi=O(\epsilon^{-2})& 0.25\epsilon^2\le x^2+y^2\le 0.5\epsilon^2 \\ \nabla\Psi=0, &x^2+y^2\le0.25\epsilon^2\end{cases}$$
with implicit constants depending on $A,\varepsilon$ and $r_c$.  We will show that by suitable choice of $A$ the condition \eqref{overlinetrchi<0} can be verified is $\delta$ is small enough for each $\varepsilon>0$. The condition \eqref{joiningcondition} will be verified by choosing $r_c$ and then deforming $\Psi$ near $\mathcal{C}$. Because we assume $T\gg \log\frac{1}{\epsilon}$,   the surface we construct  lie to the future of the initial hypersurfaces.

We assume $r_c\gg \epsilon$ and the choice of $r_c$ below will verify this condition. When $\Psi(r)=\delta$, we have $r_c-r=O(\delta)$. Therefore if $\delta$ is sufficiently small, the surface $S$ and $\underline{S}$ lie outside the waves in the intersecting region where the estimates in Proposition \ref{improvedoutsidewave} hold. The surfaces then enter the region before intersecting, where the estimates in Proposition \ref{thm:beforeintersection} hold. Also, on the whole $\Hb_\delta$ for $u\in[u_0,\delta]$, the induced metric verifies
$$|\gs-\gs_{\mathrm{Euc}}|\lesssim\delta^{\frac{1}{2}},$$
and hence
$$|\nablas\Psi|\lesssim\left|\nabla_{\gs_{\mathrm{Euc}}}\Psi\right|_{\gs_{\mathrm{Euc}}}=:|\nabla\Psi|.$$
We also have
$$|\partial_{x,y}\gs|\lesssim\delta^{\frac{1}{2}},$$
and then 
$$|\nablas^2\Psi|\lesssim|\nabla^2\Psi|+\delta^{\frac{1}{2}}|\nabla\Psi|$$
and
$$|\Deltas\Psi-\Delta\Psi|\lesssim\delta^{\frac{1}{2}}(|\nabla^2\Psi|+|\nabla\Psi|).$$

Let us go back to the condition \eqref{overlinetrchi<0}. Together with the estimates along $\Hb_\delta$ for $u\in[u_0,\delta]$, the last four terms on the left hand side is bounded by $O(\delta^{\frac{1}{2}})$  on the whole surface $\underline{S}$ where the implicit constant depends on the seed data $\psi$, $T$, $\epsilon$, $A, \varepsilon$ and $r_c$. Also, $\Deltas\Psi$ is $O(\delta^{\frac{1}{2}})$ close to $\Delta\Psi$, in order that \eqref{overlinetrchi<0} hold, we only need to check 
\begin{equation}\label{ellipticinequality}2\Delta\Psi-\Omega^{-1}\tr\chi>0.\end{equation}

In the region $0.5\epsilon^2\le x^2+y^2\le r_c^2$, we in fact have $\Omega^{-1}\tr\chi\le0$ by the Raychauduri equation (or we have $|\tr\chi|=O(\delta^{\frac{1}{2}})$). Then \eqref{ellipticinequality} follows by noting that $\Delta\Psi=\varepsilon$ and choosing $\delta$ small enough.

In the region $x^2+y^2\le 0.5\epsilon^2$, the surface lie in $\Hb_\delta$ for $u\in[u_0,0]$, so by standard argument  using Raychauduri equation and \eqref{Dbchih} (note that $\tr\chi=0$ on $\Hb_0$) we have
\begin{align*}\Omega^{-1}\tr\chi(\delta, u, x,y)\le& -\int_0^\delta |\chih|^2(\ub,u, x,y)\D\ub\\
\le& -\frac{1}{2}\int_0^\delta|\chih|^2(\ub, u_0, x(\ub,u), y(\ub, u))\D\ub\\
\le&-\frac{1}{2}M_0\epsilon^{-2}\end{align*}
if $\delta$ is sufficiently small, where $\bar{x}(\ub, u), \bar{y}(\ub, u)$ is such that the generator of  $\Hb_\delta$ starting at $(\ub, u_0, \bar{x}(\ub, u), \bar{y}(\ub, u))$ reaches $(\ub, u, x,y)$. In order that \eqref{ellipticinequality} holds, we will need
$$M_0\epsilon^{-2}> O(\epsilon^{-2}).$$ 
In fact, by a refined calculation, we have $|\Delta\Psi|\le c(A\epsilon^{-2}+\varepsilon)$ for some universal constant $c$. Therefore we need
$$M_0>c(A+\varepsilon\epsilon^2).$$
So, for any $M_0>0$ and $\epsilon\ll M_0$, we can choose $A\asymp M_0$ so that the above inequality holds and \eqref{ellipticinequality} holds. 

In order that \eqref{joiningcondition} holds, we should choose $r_c$ such that on $\mathcal{C}$
$$|\nabla \Psi|_{\gs_{\mathrm{Euc}}}=1$$
and then deform the surface $\underline{S}$ near $\mathcal{C}$. This is to say $\Psi'(r_c)=-1$ (note that $\Psi$ is a nonnegative function), so 
$$-\frac{A}{r_c}+\frac{\varepsilon r_c}{2}=-1$$
and $r_c=\frac{\sqrt{1+2\varepsilon A}-1}{\varepsilon}$. Here we may take $\varepsilon$ small enough so that $r_c\asymp A\asymp M_0$. Note that the  metric $\gs$ on $\underline{S}$ is nearly flat, the area of $\underline{S}$ is approximately $\pi r_c^2$ and the area radius of $\underline{S}$ is comparable to $M_0$. Now we take a deformation
$$\widetilde{\Psi}(r,\vartheta)=\Psi(r)+a(\vartheta)(r_c-r)\varphi\left(\frac{r_c-r}{\frac{1}{2}r_c}\right)$$
where $\vartheta$ is the angle defined by $x=r\cos\vartheta, y=r\sin\vartheta$ and $\varphi(s)$ is a smooth cutoff function  such that $\varphi(s)=1$ for $s\in[0,\frac{1}{2}]$ and $\varphi(s)=0$ for $s\ge1$. Then we have $\widetilde{\Psi}(r_c,\vartheta)=0$ and 
$$\frac{\partial\widetilde{\Psi}}{\partial r}(r_c,\vartheta)=-1-a(\vartheta),$$
$$\frac{\partial\widetilde{\Psi}}{\partial\vartheta}(r_c,\vartheta)=0.$$
So
$$\Omega|\nablas\Psi|_{\mathcal{C}}=\Omega\sqrt{\gs^{AB}(\D r)_A(\D r)_B}(1+a(\vartheta))$$
and the condition \eqref{joiningcondition} holds if we choose 
\begin{equation}\label{avartheta}a(\vartheta)=\left(\Omega(\delta,\delta,r_c,\vartheta)\sqrt{\gs^{rr}(\delta,\delta, r_c,\vartheta)}\right)^{-1}-1\end{equation}
and we also have $|a|,|a'|,|a''|\lesssim\delta^{\frac{1}{2}}$.  By direct calculation,
$$\Delta\widetilde{\Psi}-\Delta\Psi=O(\delta^{\frac{1}{2}})$$
in the cutoff region, where $\Delta\Psi=\varepsilon$, so the deformed function $\widetilde{\Psi}$ still have $\Delta\widetilde{\Psi}>0$ in the cutoff region and hence verifies \eqref{ellipticinequality}. The area radius of the deformed surface is still comparable to $M_0$.

Therefore, the surface $\underline{S}$ on $\Hb_\delta$ and $S$ on $H_\delta$, constructed as two graphs of $\widetilde{\Psi}$ on $\{x^2+y^2\le r_c^2\}\subset\Sigma_{\delta,\delta}$, are joined together along their common boundary $\mathcal{C}\subset\Sigma_{\delta,\delta}$ and we obtain a compact surface $\mathcal{S}=S\cup\underline{S}$. Its expansion relative to $\widehat{\ell}$ on $\underline{S}$ and $\widehat{\underline{\ell}}$ on $S$ is negative, with $\widehat{\ell}$ and $\widehat{\underline{\ell}}$ matching continuously across the joining curve, which is denoted by $n_+$.  Therefore, $\mathcal{S}$ is a piecewise smooth outer trapped compact surface with continuous outer null normal. 

As mentioned before, $\mathcal{S}$ itself is not smooth and the inner null normal $n_-$ of $\mathcal{S}$ is not continuous across $\mathcal{C}$. Moreover, the inner expansion of $\mathcal{S}$ may be positive somewhere. We will show that in the maximal future development, $\partial J^+(\mathcal{S})=J^+(S)\backslash I^+(S)$ is still compact so that Penrose's argument in its incompleteness theorem can still carry over to this case. 

\begin{proposition}
The maximal future development is future null geodesically incomplete.
\end{proposition}
\begin{proof}
Suppose that the maximal development is future null geodesically complete, we will show that  $\partial J^+(\mathcal{S})$ is compact. The remaining argument is then the same to the standard argument, which applies to the case when the initial data is of characteristic type (also see Prologue of Christodoulou's monograph \cite{Chr08}).

Suppose that the outer expansion of $\mathcal{S}$ verifies $\theta<\theta_0<0$, we claim that
\begin{align*}\partial J^+(\mathcal{S})\subset\bigcup_{p\in\mathcal{S}, s\in[0,-\frac{2}{\theta_0}]}\exp_{p}(sn_+)&\cup \{(\delta, \sigma_{\Lb}(u,x,y))| x^2+y^2\le r_c^2\}\\
&\cup \{(\ub, \delta ,x,y)| -\Psi(x,y)+\delta\le \ub\le\delta, x^2+y^2\le r_c^2\}\end{align*}
where $\sigma_{\Lb}(u,x,y)$ is the point on $H_u$ along the generator of $\Hb_\delta$ terminating at $(\delta, \delta, x,y)$  The first set is the part of $\partial J^+(\mathcal{S})$ consisting of future-directed outer null geodesics emanating orthogonally from $\mathcal{S}$, which is compact by standard Jacobi field argument (using admissible one-sided variations at the boundary) and the continuity of $n_+$ across $\mathcal{C}$.  The second set is the part of $\Hb_\delta$ enclosed by $\underline{S}$ and the interior of $\mathcal{C}$ and third set is the part of $H_\delta$ enclosed by $S$ and the interior of $\mathcal{C}$.  These sets are all compact and  the compactness of $\partial J^+(\mathcal{S})$ follows by its closedness (in a globally hyperbolic spacetime).
 
 The prove our claim, we pick a point $q\in \partial J^+(\mathcal{S})$, then $q$ must be connected by a null geodesic segment $\sigma$ starting at some $p\in\mathcal{S}$  without conjugate and cut points in its interior. If $p$ belongs to the smooth part of $\mathcal{S}$, then the tangent of $\sigma$ at $p$ must be orthogonal to $\mathcal{S}$. If the tangent of $\sigma$ at $p$ is $n_+$, then $q$ belongs to the first set of the union, if the tangent is $n_-$, then $q$ lies on the generators of $\Hb_\delta$ or $H_\delta$. But $q$ must belong to the last two sets of the union, since these generators reach cut points on $\Sigma_{\delta,\delta}$ in the interior of $\mathcal{C}$.  
 
 If $p\in\mathcal{C}$, and suppose that $n$ is tangent vector of  $\sigma$ at $p$. Note that the tangent (half) plane of $\underline{S}$ at $p\in \mathcal{C}$ is spanned by $\{-\partial_r+\partial_r \widetilde{\Psi}\Lb, \partial_\vartheta\}$ and the tangent (half) plane of $S$ at $p\in\mathcal{C}$ is  spanned by $\{-\partial_r+\partial_r\widetilde{\Psi} L, \partial_\vartheta\}$. Note that the plus sign before 
 $$\partial_r\widetilde{\Psi}=-1-a(\vartheta)=-(\Omega\sqrt{\gs^{rr}})^{-1}<0$$
   is chosen such that the vectors point into the interior of respective smooth piece. By first variation of arc-length argument, we must have
 \begin{align}
\label{condition1} g(n, \partial_\vartheta)=0,\\
 \label{condition2} \quad g(n,-\partial_r-(\Omega\sqrt{\gs^{rr}})^{-1}\Lb)\le0, \\
\label{condition3} \quad g(n,-\partial_r-(\Omega\sqrt{\gs^{rr}})^{-1}L)\le0.
 \end{align}
 Write $n=n^r\partial_r+n^\vartheta\partial_\vartheta+n^LL+n^{\Lb}\Lb=n^A\partial_{x^A} +n^LL+n^{\Lb}\Lb$. Because $n$ is null, we have
 $$\gs_{AB}n^An^B-4\Omega^2n^Ln^{\Lb}=0.$$
  Since $n$ is future directed, and $n^Ln^{\Lb}\ne0$ (if so, $\gs_{AB}n^An^B=0$ and $n=L$ or $\Lb$, which is impossible in view of \eqref{condition2} and \eqref{condition3} since $g(L,\Lb)=-2\Omega^2<0$), we have $n^L> 0, n^{\Lb}>0$.  From \eqref{condition1} we have $n^r\gs_{r\vartheta}+n^\vartheta\gs_{\vartheta\vartheta}=0$ so
  $$n^\vartheta=-n^r\frac{\gs_{r\vartheta}}{\gs_{\vartheta\vartheta}}$$
  and then
  $$n=n^r (\partial_r-\frac{\gs_{r\vartheta}}{\gs_{\vartheta\vartheta}}\partial_\vartheta)+n^LL+n^{\Lb}\Lb,$$
  \begin{equation}\label{nLnLb}4\Omega^2n^Ln^{\Lb}=\gs_{AB}n^An^B=\gs_{rr}(n^r)^2+2\gs_{r\vartheta}n^rn^\vartheta+\gs_{\vartheta\vartheta}(n^\vartheta)^2=(\gs^{rr})^{-1}(n^r)^2.\end{equation}
From \eqref{condition2} and \eqref{condition3}, we have
$$-n^r(\gs^{rr})^{-1}+2\Omega(\sqrt{\gs^{rr}})^{-1}n^L\le0,$$
$$-n^r(\gs^{rr})^{-1}+2\Omega(\sqrt{\gs^{rr}})^{-1}n^{\Lb}\le0.$$
and the above two inequalities imply that $n^r>0$, which means that $n$ points outward.  Together with \eqref{nLnLb}, These two inequalities imply
$$-2\Omega \sqrt{n^Ln^{\Lb}}+2\Omega n^L\le0,$$
$$-2\Omega \sqrt{n^Ln^{\Lb}}+2\Omega n^{\Lb}\le0.$$
So we have $n^L=n^{\Lb}$. We may choose $n^L=n^{\Lb}=\Omega^{-1}$,  By \eqref{nLnLb} again, 
$$n^r=2\sqrt{\gs^{rr}},$$
and then
$$n^\vartheta=-2\sqrt{\gs^{rr}}\frac{\gs_{r\vartheta}}{\gs_{\vartheta\vartheta}}.$$
So
$$n=2\sqrt{\gs^{rr}}\partial_r-2\sqrt{\gs^{rr}}\frac{\gs_{r\vartheta}}{\gs_{\vartheta\vartheta}}\partial_\vartheta+\Lh+\Lbh=2\sqrt{\gs^{rr}}\partial_r+2\left(\sqrt{\gs^{rr}}\right)^{-1} \gs^{r\vartheta}\partial_\vartheta+\Lh+\Lbh.$$
Note that 
$$\Omega\nablas_{\gs}\Psi\big|_\mathcal{C}=\Omega\partial_r\widetilde{\Psi}\gs^{rB}\partial_B=-\sqrt{\gs^{rr}}\partial_r-\left(\sqrt{\gs^{rr}}\right)^{-1}\gs^{r\vartheta}\partial_\vartheta,$$
 and we conclude that $n=n_+$ along $\mathcal{C}$.  The $q$ belongs to $\bigcup_{p\in\mathcal{S}, s\in[0,-\frac{2}{\theta_0}]}\exp_{p}(sn_+)$ and the proof is completed. Here the continuity of $n_+$ across $\mathcal{C}$ is crucial.

\end{proof}

The proof of Theorem \ref{main} is then completed.

\end{document}